\documentclass[showpacs,preprint,amssymb]{revtex4}
\usepackage{graphicx}
\usepackage{dcolumn}
\usepackage{bm}
\usepackage{pslatex}
\usepackage{amsmath}
\usepackage{amssymb}
\usepackage{pst-node}
\usepackage{pst-coil}

\begin{document}

\title{Proposal for reading out anyon qubits in non-abelian $\nu = 12/5$ quantum Hall state}

\author{Suk Bum Chung}

\affiliation{University of Illinois, Department of Physics\\ 1110 W. Green St.\\
Urbana, IL 61801 USA\\E-mail: sukchung@uiuc.edu}

\author{Michael Stone}

\affiliation{University of Illinois, Department of Physics\\ 1110 W. Green St.\\
Urbana, IL 61801 USA\\E-mail: m-stone5@uiuc.edu}

\date{\today}

\begin{abstract}
To detect non-abelian statistics in the $\nu = 12/5$ quantum Hall
state through interferometry, we apply an analysis similar to the
ones proposed for the non-abelian $\nu = 5/2$ quantum Hall state.
The result is that the amplitude of the Aharonov-Bohm oscillation of
this interference is dependent on the internal states of quasiholes,
but, in contrast to the $\nu = 5/2$ quantum Hall state, independent
of the number of quasiholes. However, if the quasiholes are in a
superposition state, it is necessary for the interferometer to have
certain additional features to obtain the coefficients.
\end{abstract}


\maketitle

\section{Introduction}

In two dimensions, particles are no longer constrained to obey Bose
or Fermi statistics. It is possible for an exchange of identical
particles to result in the multiplication of the wavefunction by an
arbitrary phase factor $e^{i\theta}$, not just -1 or +1
\cite{leinaas}. Particles with such properties are called {\it
anyons} \cite{wilczek}.

There exists an even more exotic possibility. If there is a set of
$g>1$ degenerate states $\psi_a$, $a=1,2,...,g$, for anyons with
identical configurations, exchanging particles $i$ and $j$ can
rotate one state into another in the space spanned by the
$\psi_a$'s:
\begin{equation}
\psi_a \rightarrow T^{ij}_{ab} \psi_b.
\label{EQ:stateRot}
\end{equation}
There is no reason to expect the matrices $T^{ij}_{ab}$ to commute
in general, and when they do not commute, the particles are said to
obey non-abelian braiding statistics \cite{nonAbel}.

The transformation Eq.(\ref{EQ:stateRot}) is possible because, in
two dimensions, particle exchange represents a topologically
nontrivial manipulation. External potentials or impurities, if weak
enough, cannot induce such a transformation, so the $g$-fold
degeneracy is unaffected by such perturbations. These
characteristics make non-abelian systems attractive candidates for
fault-tolerant computation \cite{freedman,kitaev_annals}. There
exist possible quantum Hall states where the braiding rules for a
set of $N$ quasiparticles coincide with the braiding rules for
$N$-point conformal blocks in the level-$k$ SU(2) Wess-Zumino-Witten
model. In the case $k=3$, these braiding rules can be used to
construct a universal quantum computer \cite{freedman}. In
particular a universal set of quantum gates realized by anyon
braiding in a $k=3$ system has been found \cite{bonesteel}.

Although it is not adequate for quantum computation, the $k=2$ case
has received more attention. This is partly because it is simpler
but also because there is experimental \cite{pan} and numerical
evidence \cite{morf,rezayi-haldane} that such a system actually
exists in the $\nu = 5/2$ quantum Hall system \cite{moore-read}.
However, the $\nu = 12/5$ quantum Hall state is a possible candidate
for the $k=3$ case, although evidence is not as strong. Consequently
it is of interest to construct a readout scheme for the $k=3$ case
similar to the $k=2$ scheme exhibited in \cite{sarma,stone-chung}.

In this paper, we will show that in the Aharonov-Bohm (AB)
oscillation experiment similar to the one proposed for the $\nu =
5/2$ quantum Hall system \cite{bonderson,stern}, the $\nu = 12/5$
quantum Hall system will show the effects of its non-abelian
statistics. The paper is organized as follows. In Section II, we
discuss the analysis of the AB oscillation experiment in the $\nu =
5/2$ quantum Hall system. In Section III, we show how to construct
the non-abelian quantum Hall state at $\nu = 12/5$. In Section IV,
we present our main result, the analysis of the AB oscillation in
the $\nu = 12/5$ quantum Hall system. Section V is a conclusion and
discussion.

\section{Detecting non-Abelian statistics in $\nu = 5/2$ quantum Hall
system}

\begin{figure}
\includegraphics[width=2.0in]{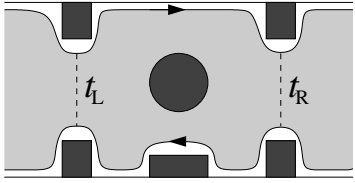}
\caption{A two point-contact interferometer for measuring the
quasiparticle statistics. The light-gray region contains an
incompressible fractional quantum Hall liquid. The front gates
(black rectangles) are used to bring the opposite edge currents
(indicated by arrows) close to each other to form two tunneling
junctions. Applying voltage to the central gate creates an antidot
in the middle and adjusts the number of quasiholes contained there.
In addition a side gate can be used in such a way that a voltage
applied to it would keep the filling fraction constant even as the
applied magnetic field is changed.} \label{Fig:AB}
\end{figure}

The basic scheme for the experiment in both filling fractions $\nu =
5/2$ and $\nu = 12/5$ is a two point-contact interferometer composed
of a quantum Hall bar with two front gates that is shown in
Fig.\ref{Fig:AB}
\cite{chamon,goldman,fradkin-nayak,sarma,stern,bonderson}. By
biasing the front gates, we can create constrictions in the Hall
bar, and so adjust the tunneling amplitude of $t_L$ and $t_R$ across
the constrictions. The tunneling between opposite-edge currents
leads to deviations of $\sigma_{xy}$ from its quantized value, or
equivalently, to the appearance of $\sigma_{xx}$. This $\sigma_{xx}$
can be measured by connecting one edge to a constant current source
and measuring the voltage drop across the other edge. For this
experiment, the interest is on the case where the tunneling
amplitudes $t_L$ and $t_R$ are small. This must be so in order to
ensure that the tunneling current is entirely due to lowest charge
($e/4$) quasiholes with no contribution from higher charge
composites \cite{bonderson}.

To the lowest order in $t_L$ and $t_R$, the tunneling current and,
hence, longitudinal conductivity $\sigma_{xx}$ in this system will
be proportional to the probability that current entering the bottom
edge leaves through the top edge \cite{bonderson,fradkin-nayak}:
\begin{eqnarray}
\sigma_{xx} &\propto& |(t_L U_L + t_R U_R)|\Psi\rangle|^2 \nonumber\\
            &=& |t_L|^2 + |t_R|^2 + 2{\rm Re}\{t^*_L t_R \langle\Psi|U^{-1}_L U_R|\Psi\rangle\} \nonumber\\
            &=& |t_L|^2 + |t_R|^2 + 2{\rm Re}\{t^*_L t_R e^{i\alpha}\langle\Psi|M_n|\Psi\rangle\}.
\label{EQ:tunnel}
\end{eqnarray}
In this expression, $U_L$ and $U_R$ are the unitary evolution
operators for a quasihole taking the two respective paths, and
$|\Psi\rangle$ is the initial state of the system. In the third
line, the operator $M_n$ is the transformation solely due to the
braiding statistics of winding a single quasihole around $n$
quasiholes \cite{fradkin-nayak}, and $e^{i\alpha}$ is the abelian
phase factor that includes the AB phase.

In this Section, we will consider $\langle\Psi|M_n|\Psi\rangle$ for
the $n$ quasiholes in the interferometer in the case $\nu = 5/2$.
The essential point is that this operator $M_n$ does not probe
individually the local properties of the quasiholes, because in this
experiment, these $n$ quasiholes are seen only as a single composite
entity. Therefore one first needs to consider what is the fusion
rule for these anyons. It is known, through the machinery of
conformal field theory, that the Moore-Read state can be built from
the ${\bm Z}_2$ Ising anyon model. This model has three particle
types, conventionally denoted as: $\mathbb{I}$ (vacuum), $\sigma$
(spin/vortex), and $\psi$ (Majorana fermion). Its non-trivial fusion
rules are:
\begin{equation}
\sigma \times \psi = \sigma, \quad \sigma \times \sigma = \mathbb{I}
+ \psi, \quad \psi \times \psi = \mathbb{I}.
\label{EQ:IsingFusion}
\end{equation}
(Note that the magnetic flux of the quasiholes does not show up in
this formalism. However, that is a less interesting issue, since for
all cases, the flux can be accounted for by some abelian phase
factors.)

Eq.(\ref{EQ:IsingFusion}) tells us that for the case $n$ even, the
composite can turn out to be either $\mathbb{I}$ or $\psi$, whereas
for $n$ odd it can only be $\sigma$. But (\ref{EQ:IsingFusion}) also
tells us that fusion of $\sigma$ and $\psi$ (not to mention $\sigma$
and $\mathbb{I}$) has a unique result. Therefore the operator $M_n$,
which is equivalent to encircling a $\sigma$, representing the
tunneling quasihole, around the composite in the island, would only
lead to wavefunction modification by some phase factor in this case.
Diagrammatically this means that a diagram where one particle winds
around another can be reduced to an unwound diagram. The phase
factor has been worked out \cite{bonderson,nayak}, and $M_n$ in this
case is effectively reduced to the diagrammatic braiding rules
worked out by Bonderson {\it et al.} \cite{bonderson}:
\begin{equation}
  \label{eq:braiding1}
  \pspicture[0.5](2.0,2.5)
  \psline[linewidth=1.4pt,linecolor=black,linestyle=dashed]
  (1,0.5)(1.0,2.5)
  \pscurve[linewidth=1.5pt,linecolor=magenta,border=4pt]
  (0.2,0.5)(1,1.0)(1.4,1.6)(1.2,1.9)
  \pscurve[linewidth=1.5pt,linecolor=magenta](0.8,2.15)(0.5,2.35)(0.2,2.5)
  \rput[bl]{0}(0,0.0){$\sigma$}
  \rput[bl]{0}(1,0){$\mathbb{I}$}
  \endpspicture
  \; =  \;
  \pspicture[0.5](2,2.5)
  \psline[linewidth=1.4pt,linecolor=black,linestyle=dashed]
  (1,0.5)(1,2.5)
  \psline[linewidth=1.5pt,linecolor=magenta](0.2,0.5)(0.2,2.5)
  \rput[bl]{0}(0,0.0){$\sigma$}
  \rput[bl]{0}(1,0){$\mathbb{I}$}
  \endpspicture
\end{equation}
\begin{equation}
  \label{eq:braiding2}
  \pspicture[0.5](2.0,2.5)
  \pscoil[coilarm=1pt,coilwidth=5pt,coilaspect=0,linewidth=1.5pt,linecolor=blue]
    (1,0.5)(1.0,2.5)
  \pscurve[linewidth=1.5pt,linecolor=magenta,border=4pt]
  (0.2,0.5)(1,1.0)(1.4,1.6)(1.2,1.9)
  \pscurve[linewidth=1.5pt,linecolor=magenta](0.8,2.15)(0.5,2.35)(0.2,2.5)
  \rput[bl]{0}(0,0.0){$\sigma$}
  \rput[bl]{0}(1,0){$\psi$}
  \endpspicture
  \; = (-1) \;
  \pspicture[0.5](2.0,2.5)
  \pscoil[coilarm=1pt,coilwidth=5pt,coilaspect=0,linewidth=1.5pt,linecolor=blue]
  (1,0.5)(1,2.5)
    \psline[linewidth=1.5pt,linecolor=magenta](0.2,0.5)(0.2,2.5)
  \rput[bl]{0}(0,0.0){$\sigma$}
  \rput[bl]{0}(1,0){$\psi$}
  \endpspicture
\end{equation}
(One way to obtain these phase factors is by examining the exponent
of the operator product expansion of these two particles.) Therefore
if one adjusts the magnetic field while maintaining the filling
fraction (something one can do by varying the size of the island
region using a side gate voltage), one will observe AB interference.

The situation is different for the case $n$ odd.
Eq.(\ref{EQ:IsingFusion}) tells us that the composite in this case
will always be $\sigma$. However, the second fusion rule of
Eq.(\ref{EQ:IsingFusion}) indicates that there are two available
states in this case. Although one can diagonalize $M_n =
(T^{tc}_{ab})^2$ for this case (where `$t$' in the superscript
represents the tunneling quasihole and `$c$' the composite particle
in the island), there are two different eigenvalues in this case
\cite{fradkin-nayak}. This means that the diagram with two
$\sigma$'s winding around each other,
\begin{equation}
  \label{eq:braiding3}
  \pspicture[0.5](2.0,2.5)
  \psline[linewidth=1.5pt,linecolor=magenta] (1,0.5)(1.0,2.5)
  \pscurve[linewidth=1.5pt,linecolor=magenta,border=4pt]
  (0.2,0.5)(1,1.0)(1.4,1.6)(1.2,1.9)
  \pscurve[linewidth=1.5pt,linecolor=magenta](0.8,2.15)(0.5,2.35)(0.2,2.5)
  \rput[bl]{0}(0,0.0){$\sigma$}
  \rput[bl]{0}(1,0){$\sigma$}
  \endpspicture
\end{equation}
cannot be reduced to the diagram of two unwound $\sigma$'s
multiplied by some phase factor. Hence, the effect of braiding
cannot be reduced to some phase factor as it was for the diagrams
(\ref{eq:braiding1}) and (\ref{eq:braiding2}). It had been found by
Bonderson {\it et al.} that the diagram (\ref{eq:braiding3}) is
actually proportional to the diagram of two unwound $\sigma$'s
exchanging a $\psi$ particle \cite{bonderson}. Their result is
consistent with Stern and Halperin's observation that for two
different tunneling quasiholes, the $M_n$'s anticommute
\cite{stern,email}. However, for the purpose of this paper, it is
sufficient to use a less general method presented below.

The interference term $\langle\Psi|M_n|\Psi\rangle$ is the
expectation value of $M_n$ evaluated at the initial state
$|\Psi\rangle$. Diagrammatically, this is represented by the
standard closure, where each worldline is looped back onto itself in
such a way as to introduce no further braiding. From
Eq.(\ref{eq:braiding3}), for $n$ odd, $\langle\Psi|M_n|\Psi\rangle$
would be equal to the following diagram
\cite{bonderson,fradkin-nayak}:
\begin{equation}
  \label{eq:link1}
  \pspicture[0.5](2.7,2.0)
  \psarc[linewidth=1.5pt,linecolor=magenta] (1.7,1.0){0.5}{180}{360}
  \psarc[linewidth=1.5pt,linecolor=magenta] (1,1.0){0.5}{0}{180}
  \psarc[linewidth=1.5pt,linecolor=magenta,border=4pt] (1,1.0){0.5}{180}{360}
  \psarc[linewidth=1.5pt,linecolor=magenta,border=4pt] (1.7,1.0){0.5}{0}{180}
  \rput[bl]{0}(0.15,0.9){$\sigma$}
  \rput[bl]{0}(2.35,0.9){$\sigma$}
  \endpspicture
\end{equation}
provided that we set loop propagators to unity. It is important to
consider the meaning of the initial state $|\Psi\rangle$. It
includes not only the internal state of the quasiholes in the island
but also the edge state as well. This is so because the operator
$M_n$ should involve creation and annihilation of the chiral
$\sigma$ field at the edge if it is to account for tunneling of a
quasihole. In this sense, the diagrammatic calculation performed
here is taking the expectation value over tunneling quasiholes as
well \cite{email}.

The anyon wordlines of the diagram (\ref{eq:link1}) are said to have
formed a {\it Hopf link}. Provided that unlinked loops are
normalized to the value of each anyon's {\it quantum dimension}, the
{\it topological S matrix} of the  anyon model can be defined in
terms of these Hopf links \cite{preskill}. If one of the anyons
forming the  Hopf link is the vacuum, the value of this Hopf link is
merely  the quantum dimension of the other anyon. In addition, the S
matrix is unitary \cite{kitaev_rev,preskill}. For the Ising model,
these results, together with the results we have from the diagrams
(\ref{eq:braiding1}) and (\ref{eq:braiding2}), enable us to obtain
all the elements of the  S matrix, and thus evaluate the diagram
(\ref{eq:link1}).

An alternative way to calculate the diagram (\ref{eq:link1}), which
we shall employ in our analysis of the $\nu = 12/5$ case, is to view
it as involving not two, but four $\sigma$ particles. (This is
possible because by Eq.(\ref{EQ:IsingFusion}), the $\sigma$ particle
can be regarded as its own antiparticle.) Consider the situation
where two pairs of $\sigma$ particles are created out of vacuum.
Then, have one $\sigma$ particle from one pair wind counterclockwise
around one $\sigma$ particle from the other pair. Finally the two
pairs are fused. The diagram (\ref{eq:link1}) is equal to the
amplitude of obtaining vacuum for both fusion processes
\cite{email}. It is known that such amplitude is zero
\cite{sarma,stone-chung,fradkin-nayak,nayak,slingerland}; indeed the
proposed NOT gate of Das Sarma {\it et al.} \cite{sarma} depends on
this result. As a consequence, there is no AB oscillation for the
case $n$ odd.

\section{Constructing ${\bm \nu = 12/5}$ quantum Hall system from ${\bm Z_3}$
parafermion}

Although the evidence is not quite as strong as that of the Pfaffian
state at $\nu=5/2$ \cite{xia}, the $k=3$ parafermion state is
considered the most likely candidate for the $\nu = 12/5$ quantum
Hall system. As conceived by Read and Rezayi \cite{read-rezayi},
$k$-cluster quantum Hall states are generalizations of the Pfaffian
state, where the wavefunction vanishes when $k+1$ electrons come
together but, in the ground state, not vanishing when $k$ or fewer
electrons come together. In this scheme, the original Pfaffian state
has $k=2$. The fusion rules of a Majorana fermion, too, can be
generalized into the fusion rule of ${\bm Z}_k$ parafermions:
\begin{equation}
\underbrace{\psi_l\times\cdots\times\psi_l}_{  k\,{\rm factors}}
 = \mathbb{I},
\label{EQ:paraFusion}
\end{equation}
where $l = 1,\ldots,k-1$.

${\bm Z}_k$ parafermions originally arose from the ${\bm Z}_k$
generalization of the two-dimensional Ising model \cite{zf}. (In
case of $k=3$, this would be equivalent to the three-state Potts
model \cite{zf,gepner}.) In such a model, there would be a ``spin"
variable $\sigma$ on each site, taking $k$ values $\omega^p$
($p=0,1,2,\ldots,k-1$), where
\begin{equation}
\omega = \exp(2\pi i/k).
\end{equation}
If the system is at a critical point, one can treat $\sigma^l$ as a
continuous conformal field $\sigma_l(x)$, $x \in {\bf R}^2$. The
critical theory retains the ${\bm Z}_k$-symmetry, so the
correlations are invariant under the transformation
\begin{equation}
\sigma_l(x) \rightarrow \omega^{ml}\sigma_l(x)
\label{EQ:ZkCharge}
\end{equation}
for arbitrary integer $m$. From Eq.(\ref{EQ:ZkCharge}), one can say
that the ${\bm Z}_k$ charge of $\sigma_l (x)$ is $l$. This model is
{\it self-dual} - that is, there exist dual conformal fields $\mu_n
(x)$ ($n=1,2,\ldots,k-1$) and all correlation functions are
invariant under the interchange $\sigma \leftrightarrow \mu$.

It is helpful to examine the $k \leq 4$ cases. The Hamiltonian on a
square lattice can be expressed as
\begin{equation}
\mathcal{H} = -J \, \sum_{\langle ij \rangle}[\sigma_i (\sigma_j)^*
+ {\rm c.c.}].
\label{EQ:Hamiltonian}
\end{equation}
Through a generalized Wannier-Kramer transformation, one can define
the dual spin variables $\mu$ on each dual lattice site. The lattice
model of Eq.(\ref{EQ:Hamiltonian}) is self-dual for these cases
\cite{savit}, which can be seen by generalizing the case of the
Ising model treated in \cite{dual}. To obtain the correlation
function $\langle \left(\mu_{j_1}\right)^n
\left(\mu_{j_2}\right)^{-n} \rangle$, we must first calculate a
transformed partition function, in which the Hamiltonian
Eq.(\ref{EQ:Hamiltonian}) is modified by
\begin{equation}
\sigma_i (\sigma_j)^* \rightarrow t_{ij} \, \sigma_i (\sigma_j)^*.
\label{EQ:disorderTrans}
\end{equation}
In Eq.(\ref{EQ:disorderTrans}), $t_{ij}=1$ if the link is not on a
path $\Gamma$ connecting $j_1$ to $j_2$. But if the link $ij$ is on
$\Gamma$, either $t_{ij} = \omega^{-n}$ or $t_{ij} = \omega^n$ (so
that if $\Gamma$ is closed, the modification results in the mapping
to an equivalent problem where all spins in sites inside gets
multiplied by $\omega^n$), hence the term `disorder' for $\mu$
variables. The correlation function $\langle
\left(\mu_{j_1}\right)^n \left(\mu_{j_2}\right)^{-n} \rangle$ is the
ratio of this transformed partition function to the original
partition function. Since $\mu^n$ can be treated as $\mu_n(x)$ when
the system is at a critical point, Eq.(\ref{EQ:disorderTrans}) tells
us that $\sigma_l (x)$ and $\mu_n (x)$ are mutually {\it semilocal}
with the mutual locality exponent of $\gamma_{\,ln} = -ln/k$. That
is, one would pick up a phase of $\exp(2i\pi\gamma_{\,ln}) =
\omega^{-ln}$ when $\sigma_l (x)$ winds counterclockwise around
$\mu_n (x)$ or vice versa.

This result can be generalized to any $k$ values \cite{zf}. A useful
analogy in thinking about the mutual locality exponent is to regard
$\sigma_l (x)$ as possessing charge $l/k$ and $\mu_n(x)$ as
possessing vortex winding $-n$; the negative of this vortex winding
is termed the dual $\tilde{{\bm Z}}_k$ charge. (A similar conclusion
can be also obtained from the $Z_N$-Villain model treated in
\cite{fradkin-kadanoff}.) Given the formulation of $\mu_n(x)$ given
here, it is natural that there exists dual $\tilde{\bm Z}_k$
invariance
\begin{equation}
\mu_n(x) \rightarrow \omega^{ns} \mu_n(x)
\end{equation}
for arbitrary integer $s$.

One can regard the ${\bm Z}_k$ parafermion field $\psi_l$ as a
holomorphic field originating from fusing $\sigma_l$ and $\mu_l$
\cite{zf}. Therefore the ${\bm Z}_k \times \tilde{\bm Z}_k$ charge
of $\psi_l$ is $(l,l)$. One result of this formulation is that the
$\psi_l$'s are not mutually local; between $\psi_l$ and $\psi_{l'}$
the mutual locality exponent is $\gamma_{\,l'l} = -2l'l/k$.
(Together with the conservation of ${\bm Z}_k \times \tilde{\bm
Z}_k$ charge, this sets both the conformal dimension and its spin at
$\Delta_l = l(k-l)/k$ \cite{zf,gepner}.) Let us now consider the
problem of constructing the electron operator from $\psi_l$. One
wants to obtain an anticommuting fermion operator with the least
possible flux attached to it \cite{read-rezayi}. In order to do so,
one should set $l=1$ and obtain
\begin{equation}
V^{\rm{para}}_{\rm{el}} = \psi_1 :\exp(i \sqrt{1 + 2/k}\,\,\varphi_c):
\label{EQ:electron}
\end{equation}
where $\varphi_c$ is a chiral free boson uniquely defined by the
two-point correlation function $\langle \varphi_c (z) \varphi_c (0)
\rangle = -\ln z$. (Note that from this point on, as we will be
dealing with quantum Hall states, we are interested only in the
holomorphic part of our conformal fields.) This sets the mutual
locality exponent of the vertex operator part (and flux attached to
one electron in units of $\Phi_0 = h/e$) at $1+2/k$, and therefore
the electron operator of Eq.(\ref{EQ:electron}) is mutually local.
(Note that for any value of $k$, this operator has the conformal
dimension of 3/2.) The wavefunction obtained by Read and Rezayi was
constructed by taking the many-point correlation function of the
electron operator of Eq.(\ref{EQ:electron}). For $k=3$ it turns out
to be \cite{read-rezayi}
\begin{equation}
\tilde{\Psi}^{(M)}_{\rm para} (z_1,\ldots,z_N) = \langle \psi_1(z_1)
\cdots \psi_1(z_N) \rangle \prod_{i<j}(z_i - z_j)^{5/3}.
\label{EQ:paraWF1}
\end{equation}

The next step is to determine the quasihole operator for $k=3$.
Before doing this we need to consider the complete fusion rules for
the ${\bm Z}_3$ Potts model. These are \cite{big_yellow_book}:
\begin{eqnarray}
\psi_i \times \psi_i &=& \psi_{3-i}, \quad \psi_i \times \psi_{3-i}
=\mathbb{I}, \nonumber\\
\psi_i \times \sigma_i &=& \varepsilon, \quad \psi_i \times
\sigma_{3-i} = \sigma_i, \quad \psi_i \times \varepsilon =
\sigma_{3-i}, \nonumber\\
\sigma_i \times \sigma_i &=& \psi_i + \sigma_{3-i}, \quad \sigma_i
\times \sigma_{3-i} = \mathbb{I} + \varepsilon, \quad
\sigma_i \times \varepsilon = \psi_{3-i} + \sigma_i
\label{EQ:fusionPotts}
\end{eqnarray}
and
\begin{equation}
\varepsilon \times \varepsilon = \mathbb{I} + \varepsilon,
\label{EQ:FiboFusion}
\end{equation}
where $i=1,2$. The $\sigma_i$'s are the primary field  of the
parafermion algebra. (Note that for $\sigma_i$, the labeling scheme
we are using here follows Read and Rezayi \cite{read-rezayi} which
is different from what is used in \cite{big_yellow_book}.) Since we
are only concerned with the holomorphic part of these fields here,
the operator product expansions (OPEs) need to be modified from
their original result in the three-state Potts model. This is
exactly analogous to the situation in the ${\bm Z}_2$ Ising model.
In this holomorphic setting, the conformal dimensions of fields are
$\Delta_{\psi} = 2/3$, $\Delta_{\sigma} = 1/15$, and
$\Delta_{\varepsilon} = 2/5$. As a result the least singular OPE
that emerges between $\psi_1$ and $\sigma_i$ or $\varepsilon$ is
\cite{read-rezayi}
\begin{equation}
\psi_1(z) \sigma_1(0) \sim {\rm const.} \frac{1}{z^{1/3}} \,
\varepsilon (0) + \cdots. \label{EQ:OPE1}
\end{equation}
There are two condition for constructing the quasihole operator. One
is that it should have the least possible charge, or equivalently,
least flux attached. The other is that it should be mutually local
with the electron operator \cite{eddy,frolich}. Eq.(\ref{EQ:OPE1})
tells us that $\sigma_1$ needs to be included to satisfy the first
condition. The vertex operator part then needs to be adjusted to
satisfy the second condition. (In order to satisfy the second
condition one needs $\Delta_{\rm{el}} + \Delta_{\rm{qh}} -
\Delta_{\lambda'}$ to be an integer, where $\lambda'$ is the fusion
product of electron and quasihole operators \cite{eddy,frolich}.)
The answer we get is
\begin{equation}
V^{\rm{para}}_{\rm{qh}} = \sigma_1 :\exp(i \sqrt{1/15}\varphi_c):.
\label{EQ:qh}
\end{equation}

Now with Eq.(\ref{EQ:qh}), the wavefunction with quasiholes can be
constructed:
\begin{eqnarray}
\tilde{\Psi}^{(M)}_{\rm para+qh} (z_1,\ldots,z_N;w_1,\ldots,w_{3n})
&=& \langle \psi_1(z_1) \cdots \psi_1(z_N) \sigma_1(w_1) \cdots
\sigma_1(w_{3n})\rangle \nonumber\\
&\times& \prod_{i<j}(w_i - w_j)^{1/15} \prod_{i,j}(z_i - w_j)^{1/3}
\prod_{i<j}(z_i - z_j)^{5/3}.
\label{EQ:paraWF2}
\end{eqnarray}
We have seen that the electron operator is mutually local to both
electrons and quasiholes. Consequently the many-electron
wavefunction in  (\ref{EQ:paraWF2}) is analytic in the electron
coordinate,  the fractional exponent of the Laughlin-like term being
canceled out by the parafermion correlation function part. The
filling fraction for this wavefunction  is determined entirely by
the Laughlin factor. This  gives $\nu = 3/5$ \cite{read-rezayi}. In
order to obtain $\nu = 2 + (1-3/5)$ from the wavefunction in
(\ref{EQ:paraWF2}), we need to fill  the first Landau level with
electrons of both spins and then apply a particle-hole
transformation to the second Landau level. From the exponent of the
$z_i - w_j$ term in Eq.(\ref{EQ:paraWF2}), one can see that the
electric charge of a quasihole is $e/5$ in the parafermion $\nu =
3/5$ quantum Hall state \cite{read-rezayi}. Since the $\nu = 12/5$
state is obtained by applying particle-hole transformation on the
$\nu = 3/5$ state, the quasiholes of Eq.(\ref{EQ:paraWF2}) becomes
excitations with the charge $-e/5$ in the parafermion $\nu = 12/5$
quantum Hall state. Conversely, there exists charge $+e/5$
excitations, such as the ones on the antidot \cite{email}, which
originate from quasielectron excitations of the $\nu = 3/5$ state.

For most of the next section, the case of the $\nu = 3/5$
parafermion state will be considered, the particle-hole inversion
being applied at the end. The inversion will result in inverting the
signs of both the quasiparticle charge and the statistical angle.

\section{Qubit measurement in $\nu = 12/5$ quantum Hall system}

As stated at the end of the last section, this section will chiefly
deal with the AB interference that would arise if
Eq.(\ref{EQ:paraWF2}) is the second Landau level {\it electron}
wavefunction. However, for $\nu = 12/5$, the wavefunction
Eq.(\ref{EQ:paraWF2}) should be that of {\it holes}, not electrons.
One can consider an analogous situation in an abelian fractional
quantum Hall state. There, the sign of the charge would be reversed
while that of the flux remains the same. As a result, the sign of
the nontrivial phase that gets accumulated when one quasihole
encircles another is reversed. There should be the same reversal of
the sign of the phase in this interference also. However it should
be noted that in this case not all phase comes from the abelian
${\rm U(1)}$ sector; reversal of sign of the charge alone cannot
explain this sign reversal. One can formulate this sign change
precisely by applying Girvin's particle-hole transformation in the
lowest Landau level \cite{ph-trans}, which include taking the
complex conjugates of the quasiparticle coordinates.

One thing to be noticed from Eq.(\ref{EQ:fusionPotts}) is that
fusions involving $\psi_i$ produce a single operator and not a sum
of operators. This indicates that the braiding $\psi_i$'s only
contribute abelian phase factors, and that in the case of braiding,
replacing $\sigma_1$ or $\sigma_2$ with $\varepsilon$ will only
result in changing the phase factor. With these consideration, the
conclusion is that all non-abelian statistics in this model can be
derived from the fusion rule Eq.(\ref{EQ:FiboFusion}) of the
$\varepsilon$ particles. This is a very important point because
Eq.(\ref{EQ:FiboFusion}) is equivalent to the fusion rule for the
Fibonacci anyons discussed by Preskill \cite{preskill}, with
$\varepsilon$ as the Fibonacci anyon. That braids of these Fibonacci
anyons can yield universal quantum computation was explicitly shown
by Bonesteel {\it et al.} \cite{bonesteel} (Fibonacci anyons may
also be realized in quantum spin systems
\cite{fendley,freedman-nayak} and rotating Bose condensates
\cite{cooper}.)

The question now is whether there is any way one can probe the
internal state of these anyons using the method explained in Section
II. In other words, we need to see if different internal states can
lead to different results for Eq.(\ref{EQ:tunnel}). Let us first
consider the system of Fibonacci anyons. Braiding in such a system
would differ from that of the $k=3$ parafermion state only by some
phase factors which we will calculate later. For this system, one
can always have either $\mathbb{I}$ or $\varepsilon$ when two or
more anyons are fused. (This fusion result is termed {\it anyonic
charge}. This charge is conserved in the braiding transformation.)
In this probe, since anyons in the interferometer are seen only as a
single entity, they form a two-state system; in this sense, they can
be considered to have formed a qubit.

If the result of fusing all the quasiholes in the interferometer is
$\mathbb{I}$, $M_n$ should be the same as it is for the Ising model
in the diagram (\ref{eq:braiding1}). Therefore
$\langle\Psi|M_n|\Psi\rangle=1$; except for some phase factors,
which we will deal with later, the AB oscillation should be the same
as the case with no quasiholes in the island.

The situation is quite different if the fusion result is
$\varepsilon$. Since in this interferometer, evaluating
$\langle\Psi|M_n|\Psi\rangle$ involves taking standard closure to
the worldlines of particles \cite{fradkin-nayak}, just as in the
case of the ${\bm Z}_2$ Ising model, the diagram to be evaluated is:
\begin{equation}
  \label{eq:FiboBraiding}
  \pspicture[0.5](2.7,2.0)
  \psarc[linewidth=1.5pt,linecolor=magenta] (1.7,1.0){0.5}{180}{360}
  \psarc[linewidth=1.5pt,linecolor=magenta] (1,1.0){0.5}{0}{180}
  \psarc[linewidth=1.5pt,linecolor=magenta,border=4pt] (1,1.0){0.5}{180}{360}
  \psarc[linewidth=1.5pt,linecolor=magenta,border=4pt] (1.7,1.0){0.5}{0}{180}
  \rput[bl]{0}(0.15,0.9){$\varepsilon$}
  \rput[bl]{0}(2.35,0.9){$\varepsilon$}
  \endpspicture
\end{equation}
(For our purpose, all propagators are set to unity, as they were in
Section II. Furthermore, all unlinked loops are normalized to
unity.)

\begin{figure}
\includegraphics[width=1.5in]{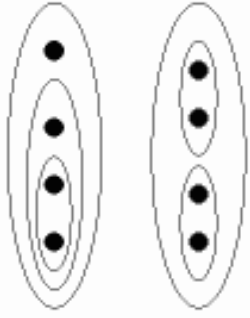}
\caption{Diagrammatic representation of two different bases for the
Hilbert space of four Fibonacci anyons. The first can be labeled by
$|((( \bullet, \bullet)_a, \bullet)_b, \bullet)_c \rangle$ and the
second $|((\bullet, \bullet)_a, (\bullet, \bullet)_b )_c \rangle$.
Here, $\bullet$ indicates one Fibonacci anyon and $a,b,c$ indicate
the fusion result of the anyons inside the bracket.}
\label{Fig:bases}
\end{figure}

From Eq.(\ref{EQ:FiboFusion}), one can see that the $\varepsilon$
particle can be regarded as its own antiparticle. Therefore, the
diagram (\ref{eq:FiboBraiding}) can be evaluated in exactly the same
way as the diagram (\ref{eq:link1}). Again one can consider the
situation where two pairs of $\varepsilon$ particles are created out
of vacuum, and one $\varepsilon$ particle from one pair is wound
counterclockwise around one $\varepsilon$ particle from the other
pair; the diagram (\ref{eq:FiboBraiding}) is equal to the amplitude
of the fusion result of both pairs being vacuum. From
Eq.(\ref{EQ:tunnel}), one can see that this amplitude would be equal
to the amplitude of the AB oscillation up to a possible phase
factor. This in turn means that for the $\nu = 12/5$ quantum Hall
state, the internal state of quasiholes in the island region
determine the amplitude of oscillation. Unlike in the case of the
$\nu = 5/2$ quantum Hall state, the amplitude of oscillation is not
determined by the number of quasiholes in the island.

Now the task is to calculate the diagram (\ref{eq:FiboBraiding}).
The elementary braid transformations for Fibonacci anyons were
derived in \cite{preskill,slingerland}, and the corresponding
transformation matrices in the three-anyon system in the basis
$\left( |((\bullet, \bullet)_{\mathbb{I}},
\bullet)_{\varepsilon}\rangle, |((\bullet, \bullet)_{\varepsilon},
\bullet)_{\varepsilon}\rangle, |((\bullet, \bullet)_{\varepsilon},
\bullet)_{\mathbb{I}}\rangle \right)$ is given in \cite{bonesteel}.
In particular the matrix for interchanging the second and third
anyons in this basis is
\begin{equation}
\sigma_2  = \left(\begin{array}{ccc} -\tau e^{-i \pi/5} & -i \sqrt{\tau}e^{-i \pi /10} & 0\\
                                     -i \sqrt{\tau}e^{-i \pi /10} & -\tau & 0\\
                                     0 & 0 & -e^{-i 2\pi/5} \end{array}\right),
\label{EQ:FiboBraid0}
\end{equation}
where $\tau = 2 \cos (2\pi/5) = (\sqrt{5} -1)/2$. The natural way to
generalize this basis to a four-anyon system would be to take the
$|((( \bullet, \bullet)_a, \bullet)_b, \bullet)_c \rangle$ basis
shown in Fig.\ref{Fig:bases}. If we are to consider the
transformation $\hat{U}$ in which the second anyon winds around the
third anyon, the last fusion result $c$ is unaffected by this
transformation, so the transformation matrix will come out in a
block-diagonal form in this basis. Therefore one obtains the
following relation:
\begin{equation}
\langle ((( \bullet, \bullet)_a, \bullet)_b, \bullet)_c|\hat{U}|((( \bullet, \bullet)_x, \bullet)_y, \bullet)_z \rangle = \delta_{cz} \, \langle (( \bullet, \bullet)_a, \bullet)_b |\hat{\sigma_2}^2|(( \bullet, \bullet)_x, \bullet)_y \rangle.
\label{EQ:threeFour}
\end{equation}
From Eq.(\ref{EQ:FiboBraid0}) and Eq.(\ref{EQ:threeFour}) one can
obtain the following matrix for this winding transformation in the
$\left( |((( \bullet, \bullet)_{\mathbb{I}}, \bullet)_{\varepsilon},
\bullet)_{\mathbb{I}} \rangle, |((( \bullet, \bullet)_{\varepsilon},
\bullet)_{\varepsilon}, \bullet)_{\mathbb{I}} \rangle, |((( \bullet,
\bullet)_{\mathbb{I}}, \bullet)_{\varepsilon},
\bullet)_{\varepsilon} \rangle, |((( \bullet,
\bullet)_{\varepsilon}, \bullet)_{\varepsilon},
\bullet)_{\varepsilon} \rangle, |((( \bullet,
\bullet)_{\varepsilon}, \bullet)_{\mathbb{I}},
\bullet)_{\varepsilon} \rangle \right)$ basis:
\begin{equation}
U = \left(\begin{array}{ccccc} -(1-\tau) & i 5^{1/4} \tau e^{-i \pi/5} & 0 & 0 & 0\\
                               i 5^{1/4} \tau e^{-i \pi/5} & -(1-\tau) e^{-i 2\pi/5} & 0 & 0 & 0\\
                               0 & 0 & -(1-\tau) & i 5^{1/4} \tau e^{-i \pi/5} & 0\\
                               0 & 0 & i 5^{1/4} \tau e^{-i \pi/5} & -(1-\tau) e^{-i 2\pi/5}  & 0\\
                               0 & 0 & 0 & 0 & e^{-i 4\pi/5} \end{array}\right).
\label{EQ:FiboBraid1}
\end{equation}

The diagram (\ref{eq:FiboBraiding}), however, needs to be calculated
in the other basis of Fig.\ref{Fig:bases}, for it is equal to
$\langle ((\bullet, \bullet)_{\mathbb{I}}, (\bullet,
\bullet)_{\mathbb{I}} )_{\mathbb{I}} | \hat{U} | ((\bullet,
\bullet)_{\mathbb{I}}, (\bullet, \bullet)_{\mathbb{I}}
)_{\mathbb{I}} \rangle$. Note that
\begin{equation}
\langle ((\bullet, \bullet)_{\mathbb{I}}, (\bullet,
\bullet)_{\mathbb{I}} )_{\mathbb{I}}|((( \bullet,\bullet)_a,
\bullet)_b, \bullet)_c \rangle = 0
\end{equation}
unless $a = \mathbb{I}$ and $c = \mathbb{I}$. Since there is only
one state in the $|((( \bullet, \bullet)_a, \bullet)_b, \bullet)_c
\rangle$ basis with $a = \mathbb{I}$ and $b = \mathbb{I}$, for some
real number $\delta$,
\begin{equation}
|((\bullet, \bullet)_{\mathbb{I}}, (\bullet, \bullet)_{\mathbb{I}}
)_{\mathbb{I}}\rangle = \exp(i\delta) ||((( \bullet,
\bullet)_{\mathbb{I}}, \bullet)_{\varepsilon},
\bullet)_{\mathbb{I}}\rangle.
\end{equation}

The amplitude for the AB oscillation when the result of fusing all
quasiholes in the interferometer is $\varepsilon$ can be given now:
\begin{eqnarray}
  \label{EQ:ampFibo}
  \langle\Psi|M_n|\Psi\rangle &=&
  \pspicture[0.5](2.7,2.0)
  \psarc[linewidth=1.5pt,linecolor=magenta] (1.7,1.0){0.5}{180}{360}
  \psarc[linewidth=1.5pt,linecolor=magenta] (1,1.0){0.5}{0}{180}
  \psarc[linewidth=1.5pt,linecolor=magenta,border=4pt] (1,1.0){0.5}{180}{360}
  \psarc[linewidth=1.5pt,linecolor=magenta,border=4pt] (1.7,1.0){0.5}{0}{180}
  \rput[bl]{0}(0.15,0.9){$\varepsilon$}
  \rput[bl]{0}(2.35,0.9){$\varepsilon$}
  \endpspicture
  \nonumber\\
  &=& \langle ((\bullet, \bullet)_{\mathbb{I}}, (\bullet, \bullet)_{\mathbb{I}} )_{\mathbb{I}} | \hat{U} | ((\bullet, \bullet)_{\mathbb{I}}, (\bullet, \bullet)_{\mathbb{I}} )_{\mathbb{I}} \rangle \nonumber\\
  &=& \langle ((( \bullet, \bullet)_{\mathbb{I}}, \bullet)_{\varepsilon}, \bullet)_{\mathbb{I}} | \hat{U} | ((( \bullet, \bullet)_{\mathbb{I}}, \bullet)_{\varepsilon}, \bullet)_{\mathbb{I}} \rangle \nonumber\\
  &=& -(1-\tau) = -\frac{3-\sqrt{5}}{2} \approx -0.382.
\end{eqnarray}
Note that there is also the phase factor of -1.

(From Preskill \cite{preskill}, one can easily obtain the S matrix
for the Fibonacci anyon model. This is possible because the braiding
involving the vacuum is trivial and the S matrix here is a $2 \times
2$ unitary matrix. The S matrix obtained in this way agrees with the
amplitude of AB oscillation in Eq.(\ref{EQ:ampFibo}).)

Some phase factors were lost by identifying the $\sigma_i$'s with
$\varepsilon$ and the $\psi_i$'s with $\mathbb{I}$. Two points need
to be made in order to figure these out. First, from
Eq.(\ref{EQ:fusionPotts}), $\sigma_1$ can be regarded as resulting
from fusing the Fibonacci anyon $\varepsilon$ and the parafermion
$\psi_2$: the quasihole tunneling can be regarded as tunneling of
the composite of the parafermion $\psi_2$ and the Fibonacci anyon
$\varepsilon$. Second, from the operator product expansion
\begin{equation}
\psi_i (z) \varepsilon (0) \sim {\rm const.} \, \frac{1}{z} \, \sigma_{3-i} (0),
\end{equation}
one can see that the $\psi_i$'s are relatively local to
$\varepsilon$ - that is, there is no accumulation of nontrivial
phase factor when $\psi_i$ winds around $\varepsilon$. Therefore the
phase factors that need to be calculated comes entirely from
$\psi_i$'s. In other words, the phase factors to be considered now
originate from the ${\bm Z}_3 \times \tilde{\bm Z}_3$ charge of the
parafermion.

If the number of quasiholes in the island is $3n+1$, the fusion result is
\begin{equation}
\underbrace{\sigma_1\times\cdots\times\sigma_1}_{  3n+1\,{\rm factors}}
 = \psi_2 \times (\mathbb{I}+ \varepsilon).
\label{EQ:paraState1}
\end{equation}
(If $n=0$ there cannot be a fusion result of $\psi_2$, of course.)
So the effect that had been ignored is that of encircling $\psi_2$
counterclockwise around each other. From Section III. this phase
factor can be found from the mutual locality exponent - the phase
factor is $\exp(2\pi i/3)$.

Similarly,
\begin{equation}
\underbrace{\sigma_1\times\cdots\times\sigma_1}_{  3n+2\,{\rm factors}}
 = \psi_1 \times (\mathbb{I}+ \varepsilon).
\label{EQ:paraState2}
\end{equation}
Here the phase factor is the same as the one that arises when
$\psi_1$ circles counterclockwise around $\psi_2$, or vice versa,
which turns out to be $\exp(-2\pi i/3)$.

On the other hand,
\begin{equation}
\underbrace{\sigma_1\times\cdots\times\sigma_1}_{  3n\,{\rm factors}}
 = \mathbb{I}+ \varepsilon.
\label{EQ:paraState3}
\end{equation}
gives rise to no further phase factor.

This shows that whereas the number of quasiholes in the
interferometer does not determine the amplitude of the oscillation,
it does induce phase shift in the oscillation. The phase shift due
to the electric charge and magnetic flux of quasiholes now needs to
be accounted for. For the quantum Hall state of
Eq.(\ref{EQ:paraWF2}), the quasihole has charge of $e/5$ and flux of
$\Phi_0/3$, the flux phase factor that arises when one quasihole is
added is $2\pi/15$. Combining these two phase shifts, we see that
there is a $2\pi / 3 + 2\pi / 15 = 4\pi/5$ phase shift per one
quasihole.

We now have the result of Eq.(\ref{EQ:tunnel}) for the $\nu = 12/5$
parafermion quantum Hall state, keeping in mind, however, that the
particle-hole inversion gives a negative sign to these phase shifts
due to the quasiholes. In this case the fusion result of $n$
quasiholes in the island region would have the three-state Potts
label of $\mathbb{I}$, $\psi_1$ or $\psi_2$:
\begin{equation}
\sigma_{xx} \propto |t_1|^2 + |t_2|^2 + 2 |t_1||t_2| \cos
\left[\alpha + {\rm arg}(t_2/t_1) - n \frac{4 \pi}{5}\right].
\label{EQ:result1}
\end{equation}
Otherwise the three-state Potts label of the fusion result would be
$\varepsilon$, $\sigma_2$ or $\sigma_1$ and one would have
\begin{equation}
\sigma_{xx} \propto |t_1|^2 + |t_2|^2 - 2 (1 - \tau) |t_1||t_2| \cos
\left[\alpha + {\rm arg}(t_2/t_1) - n \frac{4 \pi}{5}\right].
\label{EQ:result2}
\end{equation}
The first two phase terms, $\alpha + {\rm arg}(t_2/t_1)$, can be
varied by changing $B$. Since the quasiparticle charge is $e/5$, the
period of AB oscillation is $5\Phi_0$.

As far as the phase of the oscillation is concerned,
Eqs.(\ref{EQ:result1}) and (\ref{EQ:result2}) give the same result
as the hierarchial $\nu=12/5$ quantum Hall state; the non-abelian
nature of the parafermion quantum Hall state is manifest only
through the changed amplitude of the oscillation in
Eq.(\ref{EQ:result2}).

\section{Discussion}

As analyzed above, comparing the AB oscillation in the $\nu =$ 5/2
and 12/5 quantum Hall states shows that different fusion rules leads
to qualitatively different results. The biggest difference is that,
unlike the $\nu = 5/2$ case, there is no instance in the $\nu =
12/5$ case where the interference vanishes due to the number of
quasiholes in the island region. In fact, in the case of $\nu =
12/5$ for any number of quasiholes in the interferometer, the change
in the amplitude of oscillation due to the internal state of the
quasiholes occur in the same manner.

This paper does not present a  complete readout scheme for the
internal state of the quasiholes of the $\nu = 12/5$ quantum Hall
state to the extent done in the $p_x + ip_y$ superconductor
\cite{stone-chung}. Given a set of quasiholes, their total anyonic
charge is conserved in any topological process. This means that,  in
the case that the quasiholes inside the interferometer are not in a
state with  definite total anyonic charge, it is not possible to
obtain some of the phase relations between coefficients in the
superposition state. Even if the {\it total\/}  anyonic charge is
fixed, as we saw in Section IV, once  there are three or more
quasiholes, they can  be in a linear superposition of more than one
internal state. It is  impossible to probe the internal Hilbert
space of such  a quasihole cluster unless we  can move the
quasiholes adiabatically out of the interferometer region. In
addition, if we are  to obtain any phase relation between the
coefficients, we must be able to braid quasiholes adiabatically.
Without introducing such additional features, we  cannot extract the
coefficients of the internal quasihole superposition state from this
interference experiment.

Lastly it should be noted that for these ``qubits", the
initialization process is not known. So far, it is not clear how one
can prepare  quantum mechanically pure states; more work needs to be
done in this direction.

\section*{Note Added}

While this paper was in preparation, the authors learned about a
similar work by P.~Bonderson, K.~Shtengel, and J.~K.~Slingerland
\cite{race_winner}. They demonstrated that the monodromy matrix
element can be written in terms of the S matrix. By obtaining S
matrix for the general ${\bm Z}_k$ parafermion theory, they obtained
the same conclusion on the phase and amplitude of AB oscillation
presented in this paper.

\section{Acknowledgement}

S.C. would like to thank N.~Bonesteel for explaining the connection
between Fibonacci anyons and their connection to ${\rm SU}(2)_3$
algebra, P.~Bonderson and K.~Shtengel for both illuminating
explanation of the result of their previous paper \cite{bonderson},
and their comments on this paper, and B.~I.~Halperin and S.~H.~Simon
for their discussion on the issue of the sign of quasiparticle
charge in the two-point contact quantum Hall interferometry. This
work is partly funded by the University of Illinois at
Urbana-Champaign Research Board.

\end{document}